\documentclass[twocolumn,english,prl,showpacs,preprintnumbers,superscriptaddress]{revtex4-1}
\usepackage[utf8]{inputenc}
\usepackage{color}
\usepackage{amsmath}
\usepackage{amssymb}
\usepackage{graphicx}
\usepackage{textcomp}
\usepackage{dcolumn}
\usepackage{bm}
\usepackage[right]{eurosym}
\usepackage{float}
\usepackage[english]{babel}
\usepackage{blindtext}
\usepackage{babel}

\begin{document}

\title{Fano Resonances observed in Helium Nanodroplets}

\author{A. C. LaForge}
\email{aaron.laforge@physik.uni-freiburg.de}
\author{D. Regina}
\affiliation{Physikalisches Institut, Universit{\"a}t Freiburg, 79104 Freiburg, Germany}
\author{G. Jabbari}
\author{K. Gokhberg}
\author{N. V. Kryzhevoi}
\affiliation{Physikalisch-Chemisches Institut, Universit{\"a}t Heidelberg, 69120 Heidelberg, Germany}
\author{S. R. Krishnan}
\affiliation{Department of Physics, Indian Institute of Technology - Madras, Chennai 600 036, India}
\author{M. Hess}
\affiliation{Physikalisches Institut, Universit{\"a}t Freiburg, 79104 Freiburg, Germany}
\author{P. O'Keeffe}
\author{A. Ciavardini}
\affiliation{CNR - Istituto di Struttura della Materia, CP10, 00016 Monterotondo Scalo, Italy}
\author{K. C. Prince}
\author{R. Richter}
\affiliation{Elettra-Sincrotrone Trieste, 34149 Basovizza, Trieste, Italy}
\author{F. Stienkemeier}
\affiliation{Physikalisches Institut, Universit{\"a}t Freiburg, 79104 Freiburg, Germany}
\author{L. S. Cederbaum}
\affiliation{Physikalisch-Chemisches Institut, Universit{\"a}t Heidelberg, 69120 Heidelberg, Germany}
\author{T. Pfeifer}
\author{R. Moshammer}
\affiliation{Max-Planck-Institut f{\"u}r Kernphysik, 69117 Heidelberg, Germany}
\author{M. Mudrich}
\affiliation{Physikalisches Institut, Universit{\"a}t Freiburg, 79104 Freiburg, Germany}

\begin{abstract}
Doubly-excited Rydberg states of He nanodroplets have been studied using synchrotron radiation. We observed Fano resonances related to the atomic 2,0$_n$ series as a function of droplet size. Although qualitatively similar to their atomic counterparts, the resonance lines are broader and exhibit a shift in energy which increases for the higher excited states. Furthermore, additional resonances are observed which are not seen in atomic systems. We discuss these features in terms of localized atomic states perturbed by the surrounding He atoms and compare them to singly excited droplets. 
\end{abstract}

\date{\today}

\maketitle

Electronic correlation is of fundamental importance in atomic and molecular systems. Especially the interaction of energetic photons with multiple electron systems is governed by electron correlation leading to processes such as shake-off in single photon double ionization~\cite{Pattard2003}, post collision interaction in Auger processes~\cite{Russek1986}, and doubly-excited autoionization~\cite{Madden1963}. Furthermore, the presence of weakly-bound neighboring atoms offers additional correlative processes between atoms such as interatomic Coulombic decay (ICD)~\cite{Cederbaum1997} and has been fertile research subject in recent years. In particular for the He dimer, electronic correlation via ICD was shown to exist even though the He atoms have a separation of 52 \AA\, in the ground state~\cite{Sisourat2010,Havermeier2010}.

Generally speaking, electronically excited free atoms and small molecules exhibit a wealth of fine structure in their Rydberg and double excitation spectra, while in condensed matter these structures become broad and less resolved or simply vanish. The general explanation for this effect is that the wave functions of excited atoms are spatially extended, and in the condensed state, these wavefunctions are strongly perturbed by their neighbors. Confinement of the final state orbital may lead to energy shifts, and singly excited Rydberg states may evolve into excitons when condensed (e.g.~\cite{Woermer1991,Joppien1993}). 

Doubly excited states occupy a special place in studies of electronic structure as their existence is due to electron correlation. As they contain two excited electrons, they are more sensitive to perturbation by the environment. As originally shown by Fano~\cite{Fano1961}, the ionization cross section is given by
\begin{align}
	\centering
	\sigma_{Fano} = \sigma_{0} \frac{(q + \epsilon)^2}{(1 + \epsilon^2)}\mathrm{,}\,\,\mathrm{where}\,\, \epsilon = \frac{E - E_0}{\hbar(\Gamma/2)}
\end{align} 
is the reduced energy containing  $E_0$  and $\Gamma$ as the respective position and width of the resonance, $\hbar$ is Planck’s constant, and $\sigma_0$ is the ionization cross section in absence of the double excitation resonance. At the double excitation resonance, the ionization cross section exhibits an interference between the direct ionization and autoionization pathways. For an overview of the work on this topic, see Rost et al.~\cite{Rost1997}.


For rare gas clusters and condensed neon, broad features were observed near the well-known atomic window resonances~\cite{Madden1963} attributed to surface and bulk excitations, but the assignments of the features has remained tentative~\cite{Mueller1993,Thissen1998,Pavlychev2000,Zhang2009,Kassuehlke2012}. As these studies were performed on crystalline rare gas systems, the broadening may be inhomogeneities due to site-specific interactions, rather than due to the heterogeneous environment of the excitation. He droplets offer a unique system of an extremely weakly bound van der Waals superfluid with a homogenous density distribution. Besides, as the helium atom is the simplest system to study double excitations, it is therefore amenable to accurate calculation~\cite{Burgers1995}.

In this Letter, we report the resonant double excitation of He atoms condensed into nanodroplets, measured by XUV synchrotron radiation. In contrast to a recent series of experiments where many atoms in the droplet were resonantly excited by intense XUV radiation~\cite{LaForge2014,Ovcharenko2014}, here, the excitation is limited to a single atom within the droplet and Fano profiles similar to those of atomic systems are observed when scanning across the resonances. The peak positions and widths along with the resonance energies are discussed in the context of perturbation induced by couplings to the cluster.

\begin{figure}
\begin{center}{
\includegraphics[width=0.5\textwidth]{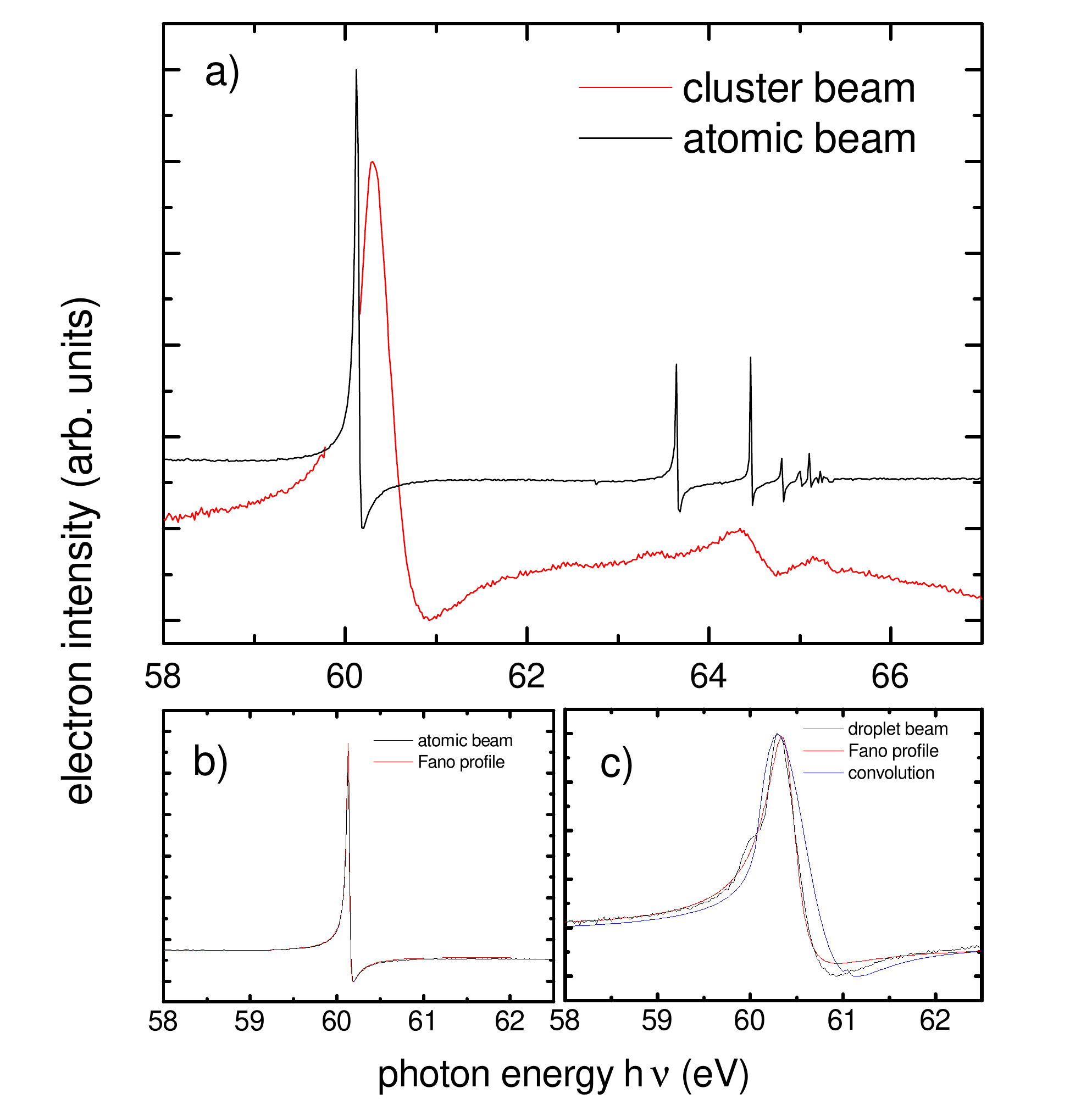}}
\caption{a) Electron signals of He atoms at a nozzle temperature of T\,=\,40 K (black) and of He droplets (red) consisting of 10$^9$ He atoms (T = 8 K) (red) as a function of the photon energy. b) Fano profile fit (red) to the 2,0$_2$ resonance of the atomic state (black) c) Fano profile fit (red) to the 2,0$_2$ resonance of the droplet state (black) The atomic 2,0$_{2}$ resonance convoluted with the line profile of the droplet 1s2p singly excited state taken from Ref.~\cite{Joppien1993} (blue).}
\label{fig1}
\end{center}
\end{figure}

The experiment was performed using a mobile He droplet source attached to an imaging photoelectron-photoion coincidence (PEPICO) detector at the GasPhase beamline of Elettra-Sincrotrone Trieste, Italy. The setup has been described in some detail earlier~\cite{OKeeffe2011,Buchta2013}, and only the significant parts will be addressed here. A beam of He nanodroplets is produced by continuously expanding pressurized He (50 bar) of high purity (He 6.0) out of a cold nozzle (T = 7 $-$ 40 K) of diameter of 5 $\mu$m into a UHV chamber. Under these expansion conditions, the mean droplet sizes range from 1 to 10$^{11}$ He atoms per droplet~\cite{Toennies2004,Stienkemeier2006}. After passing a skimmer (0.4 mm) and a mechanical beam chopper for discriminating the droplet beam signal from the He background, the He droplet beam next crosses the synchrotron beam inside a PEPICO detector consisting of an ion time of flight detector and a velocity map imaging detector operating in coincidence. For the experiments reported in this work, only the electron and mass-gated ion signals were measured for photon energies below the doubly excited resonances ($h\nu$ = 58 $-$ 68 eV)~\cite{Madden1963}. The photon energies of the gap and monochromator were scanned simultaneously with a typical step size of 20 meV and energy resolution of $E/\Delta E$ of $\approx$ 10$^4$. The peak intensity in the interaction region was estimated to be around 15 W/m$^{2}$. The intensity of the radiation was monitored by a calibrated photodiode and all photon-energy-dependent ion and electron spectra shown in this work are normalized to this intensity.  

Fig.\,\ref{fig1}\,a) shows the electron signals from a He beam at a nozzle temperature of 40 K (black line) which is almost exclusively atomic (\textgreater 99 \%), and at a nozzle temperature of 8 K (red line) which forms a droplet beam with an average size of 10$^9$ atoms. For the atomic signal, we observe the same resonance lines as were initially reported by Madden et al.~\cite{Madden1963}. The relative peak heights of the resonances from higher quantum numbers are slightly more imprecise due to their narrow widths and the finite scan steps chosen for this measurement. As these resonances have been measured numerous times (e.g.~\cite{Domke1996}), here we will only use their well-known resonance energies and shapes as references for the droplet resonances. In comparison, the resonances observed in the He droplet case are significantly broadened and blue-shifted. In addition to the electron signal, ion signals correlated to the He monomer and dimer showed a similar lineshift and broadening. Due to its close proximity to the atomic resonance, we attribute the large feature around $h\nu$ = 60.4 eV to the droplet equivalent of the 2,0$_2$ resonance. To gain a more quantitative understanding of the droplet resonances, we fit the resonance with Fano's formula for the cross-section, Eq. 1. The results are compared to the experimental spectra in Fig.\,\ref{fig1}\,b) and c) for both the atomic and droplet resonance, respectively. From the fit, we find the resonance energy and width for the droplet to be 60.4 eV and 420 meV, respectively while the $q$ parameter is -2.36. In comparison, the resonance energy and width for the equivalent atomic resonance are 60.15 and 37 meV, respectively~\cite{Domke1996}. Thus, the droplet resonance is blue-shifted by about 300 meV while the $q$ parameter is  slightly lower than that of the atomic state, $q = -2.77$.

Similar line broadening and energy shifts were previously observed for single excitations of He droplets by Joppien et al.~\cite{Joppien1993}. In that work, fluorescence spectra of He clusters were measured for photon energies of $h\nu$ = 20 - 25 eV. The observed band structures were compared to absorption lines of heavier rare gas clusters~\cite{Schwentner1985} which were explained in terms of either the Frenkel or Wannier exciton model. However, exciton models could not explain the results observed for He clusters. The interpretation given by Joppien et al.~\cite{Joppien1993} in terms of perturbed localized excited atoms instead of excitons as in heavier rare gas clusters is mainly based on the low density as well as the low dielectric constant of liquid He. In this model it is observed that for the lowest singly excited states, the exciton radius is smaller than the nearest-neighbor distance of He atoms in liquid droplets. The blue-shift is then interpreted as due to the repulsive interaction between the excited electron and the neighboring He environment whereas the large linewidth is related to both the inhomogeneous density in the extended surface region of the cluster as well as quantum mechanical density fluctuations. Recently, Kornilov et al.~\cite{Kornilov2011} used a simple model which gave good agreement with experiment to explain these effects by treating the excitation as localized at a single atom within the droplet which is then perturbed by the mean field of the surrounding He atoms. Plotted in Fig.\,\ref{fig1}c)  is the atomic 2,0$_{2}$ resonance convoluted with the line profile of the 1s2p singly excited state of the droplet from Joppien et al.~\cite{Joppien1993} together with the droplet 2,0$_{2}$ resonance for a nozzle temperature of 8 K. Overall, there is very good agreement between the convoluted atomic data and the droplet data and, therefore we believe the observed line broadening and lineshifts in doubly excited He droplets can be attributed to similar interactions between localized excited atoms and the surrounding ground state He atoms, as in singly excited droplets. 

\begin{figure}
\begin{center}{
\includegraphics[width=0.5\textwidth]{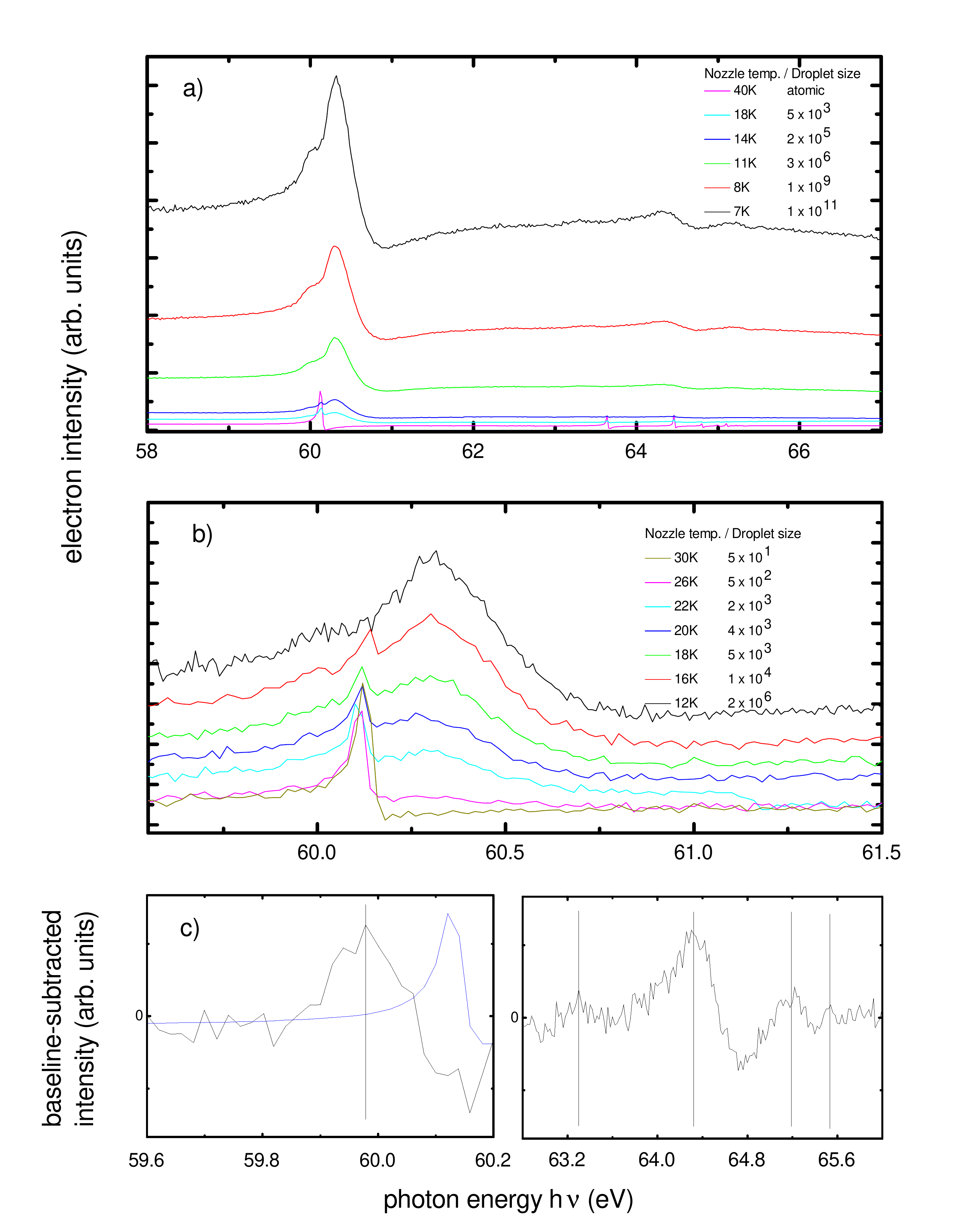}}
\caption{a) Electron signals of He at various nozzle temperatures corresponding to gas phase atoms to nanodroplets of 10$^{11}$ atoms as a function of the photon energy. b) electron spectra centered around the 2,0$_2$ resonance. c) Baseline-subtracted electron signal for the energy ranges of 58.8 - 60.2 eV (left) and 61 eV - 66 eV (right) along with the atomic 2,0$_{2}$ resonance (blue line).
}
\label{fig2}
\end{center}
\end{figure}

In Fig.\,\ref{fig2}\,a), the electron signals are plotted for droplet sizes ranging from $\langle$N$\rangle$ $\approx$ 1 $\cdot$ 10$^{11}$ atoms in the full range of measured photon energies. In contrast to atomic resonances which feature vanishing signal at the minima of the Fano profile, the droplet resonances show a large background signal on which the profile resides. Additionally, for increasing droplet size, the total electron intensity increases along with the contrast (maxima to minima in the interference) of the droplet Fano resonances. For the atomic case, the decay mechanisms are limited to either autoionization or the slower fluorescence decay~\cite{Penent2001} resulting in nearly perfect contrast in the interference between direct ionization and autoionization. Droplets, on the other hand, exhibit a much weaker interference contrast due to the presence of additional decay mechanisms in this condensed system. Besides the aforementioned autoionization and fluorescence decay, ICD is energetically allowed~\cite{Havermeier2010} as well as direct ionization of the dimer, trimer, and higher oligomer states~\cite{Peterka2007}. 

The next question which arises is whether the contributions of the homogeneous vibronic coupling versus inhomogeneous effects related to the varying density at the droplet surface? Fig.\,\ref{fig2}\,b) shows the 2,0$_{2}$ resonance for the transition from single He atoms to clusters. For small clusters, $\langle$N$\rangle$ $\leq$ 500 atoms at T $\geq$ 26 K, the features of the Fano profiles are nearly identical (in terms of line shape and width) with their atomic counterparts. This can be explained by the large atomic contribution to the supersonic gas jet. At $\langle$N$\rangle$ $\approx$ 2,000 (T = 22 K), the broadened Fano profile initially appears with nearly identical parameters ($E_0$ = 60.39 eV, $\Gamma$ = 400 meV, $q$ = -2.2) to those of the larger droplets in Fig.\,\ref{fig1} when the atomic profile is subtracted.  As the average size of the nanodroplets increases, the droplet Fano profile becomes the dominant feature observed in the spectra superseding the atomic profile at T = 12 K. The surprising observation that the shape of the resonance profile remains nearly constant irrespective of the starkly changing ratio of surface to bulk atoms for the various droplet sizes indicates that inhomogeneous broadening is negligible. On the contrary, the observed profile can be regarded as the characteristic line shape of bulk superfluid He which is dominated by homogeneous broadening due to the repulsive interaction of excited atoms with the surrounding He. Overall, these results slightly differ from previous results on singly excited He clusters where for small He clusters von Haeften et al.~\cite{Haeften2005} observed a steady broadening of the atomic lines with increase in cluster size. However, the discrepancies between the two results could possibly be explained by a larger atomic contribution to the cluster beam in the previous experiment due to the nozzle being located close to the interaction region whereas the cluster beam described here traverses two differentially pumped vacuum chambers thus resulting in a lower atomic contribution to the cluster beam.

\begin{figure}
\begin{center}{
\includegraphics[width=0.5\textwidth]{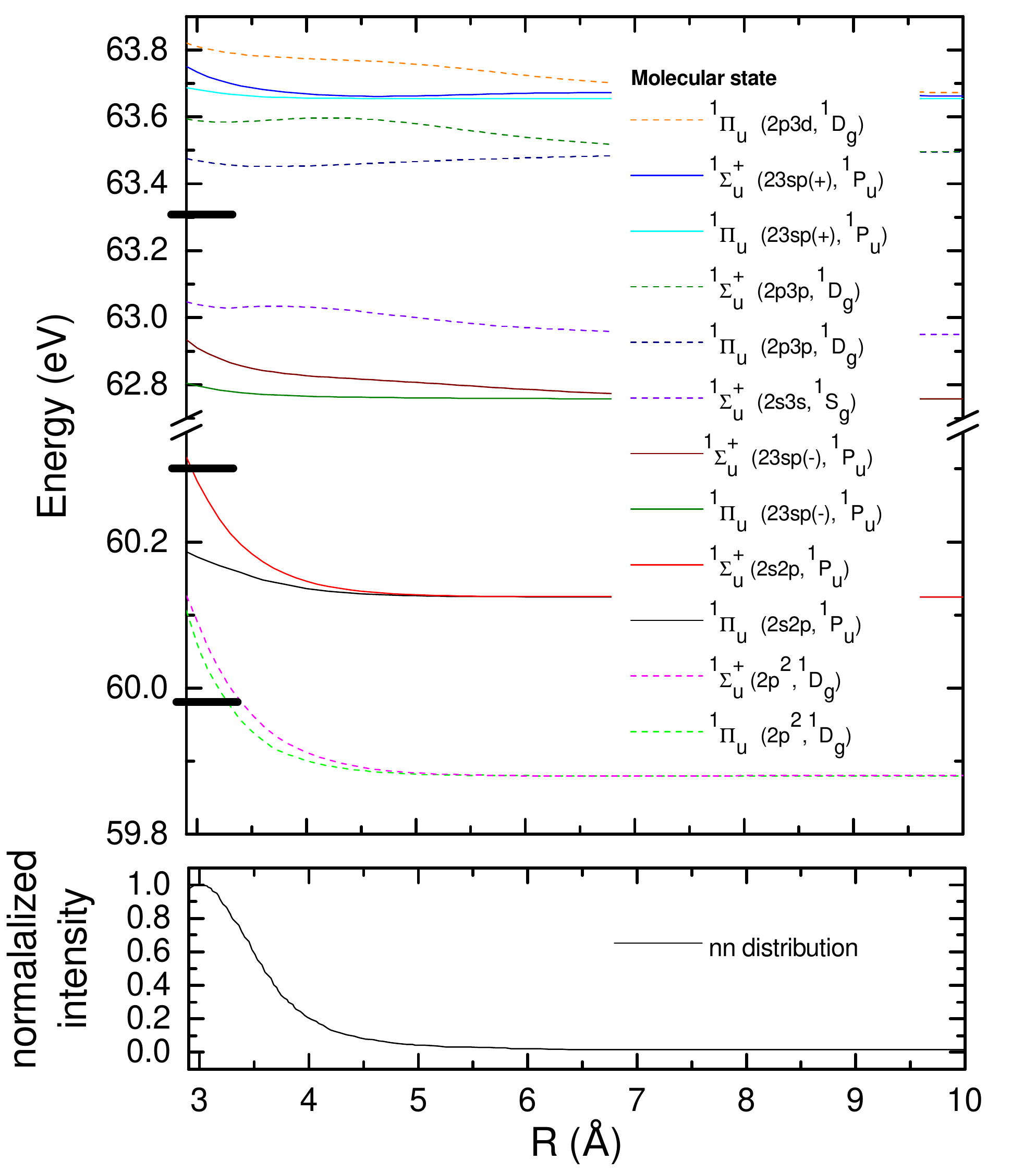}}
\caption{Computed potential energy curves for 2,0$_2$ (middle) and 2,0$_3$ (top) resonances in He dimers along with the nearest neighbor (nn) distribution (bottom) for He droplets (taken from ~\cite{Peterka2007}). The potential curves accessible by dipole-allowed transitions are plotted as solid lines, the dipole-forbidden ones as dashed lines. The first three droplet resonances are given as tick marks on the y axis.}
\label{fig3}
\end{center}
\end{figure}

For droplet sizes of $\langle$N$\rangle$ $\geq$ 3\,$\cdot$\,10$^7$ atoms (T = 11 K) and greater, additional Fano profiles appear at higher photon energies in Fig.\,\ref{fig2}\,a). To isolate these profiles as well as any additional features, we fit the background signal with a low-order polynomial, and perform a baseline subtraction for a droplet size of 10$^{11}$ atoms. Fig.\,\ref{fig2}\,c) shows the baseline-subtracted electron intensity for the photon energy ranges of 58.8 eV - 60.2 eV (left) and 61 eV - 66 eV (right), respectively. Fig.\,\ref{fig2}\,c) (left) additionally includes the atomic 2,0$_{2}$ resonance (blue line) due to its close proximity to the possible droplet resonance. Black vertical lines indicate the energies of possible resonances. The most prominent features are the three sequential resonances at 64.3 eV, 65.3 eV, and 65.5 eV (peak position). These are assigned to the higher members of the 2,0$_{n}$ series. Compared to the blueshift observed in singly excited droplets~\cite{Joppien1993} and the 2,0$_{2}$ droplet resonance, the states with higher quantum numbers exhibit a much larger shift in energy (about 720 (40) meV for the droplet 2,0$_{3}$ resonance and about 815 (50) meV for the droplet 2,0$_{4,5}$ resonance of a droplet consisting of 10$^{11}$ atoms). Since the blueshift is due to the repulsive interaction between the excited electrons and the surrounding He environment, the larger blueshift for higher quantum numbers is due to a larger orbital radius which then feels a stronger repulsion from the surrounding He. There are two additional peaks at about 60.0 eV and 63.3 eV seen in Fig.\,\ref{fig2}\,c). Given the symmetry breaking which occurs at the droplet surface, additional dipole forbidden atomic resonances are allowed in the droplet. To exclude the possibility that the resonance at 60.0 eV is simply due to the subtraction of the atomic resonance at 60.1 eV, the atomic resonance is included in Fig.\,\ref{fig2}\,c) (left).

In general, fully quantum mechanical theoretical calculations on large scale systems such as nanodroplets are challenging. Instead, ab initio calculations on simpler systems (e.g. dimers) can give insight into the effect of complex formation on the electronic properties of a doubly excited atom. Therefore, we calculated the potential energy curves for He dimers for the lower doubly excited states which are shown in Fig.\,\ref{fig3} along with the nearest neighbor distribution (bottom) for He droplets~\cite{Peterka2007}. For details of the theoretical calculation of the potential energy curves, see the supplementary material. For comparison, the peak positions of the first three droplet resonances are given as tick marks on the vertical scale. Overall, the n = 2 droplet resonances are in very good agreement with the theoretical potential energy curves if one considers vertical transitions at He-He nearest neighbor distances ($\sim$\,3~\AA) inside the droplet. The peak at 60.0 eV corresponds most likely to the dipole forbidden 2p$^2$ state. In fact, the blueshift would fit better to an average distance of about $\sim$\,3.6~\AA~corresponding to a surface state~\cite{Peterka2007}. The 2s2p state is further blueshifted than the 2p$^2$ state but is also in good agreement with the molecular potential curves. Furthermore, the splitting of the potential curves along with the broad nn distribution could therefore partially explain the observed broadening and blueshifting. For the n = 3 states the droplet resonances exhibit a much stronger blueshift than given by the molecular states. For instance the droplet 2,0$_{3}$ resonance ($h\nu$ = 64.3 eV) corresponding to the blue and cyan lines falls outside of the range of Fig.\,\ref{fig3}. Assuming a similar blueshift ($\sim$\,500~meV) for the resonance at 63.3 eV, one could assign it to the atomic 2,0$_{3}$- resonance~\cite{Domke1996} or with a slightly smaller blueshift ($\sim$\,500~meV) to the dipole forbidden 2s3s state ~\cite{Childers2006}. Overall, the experimental data is in good agreement with the theoretical calculations especially considering that the calculations were performed for He dimers whereas the atoms are an average six-fold coordinated in liquid He.

\begin{figure}
\begin{center}{
\includegraphics[width=0.5\textwidth]{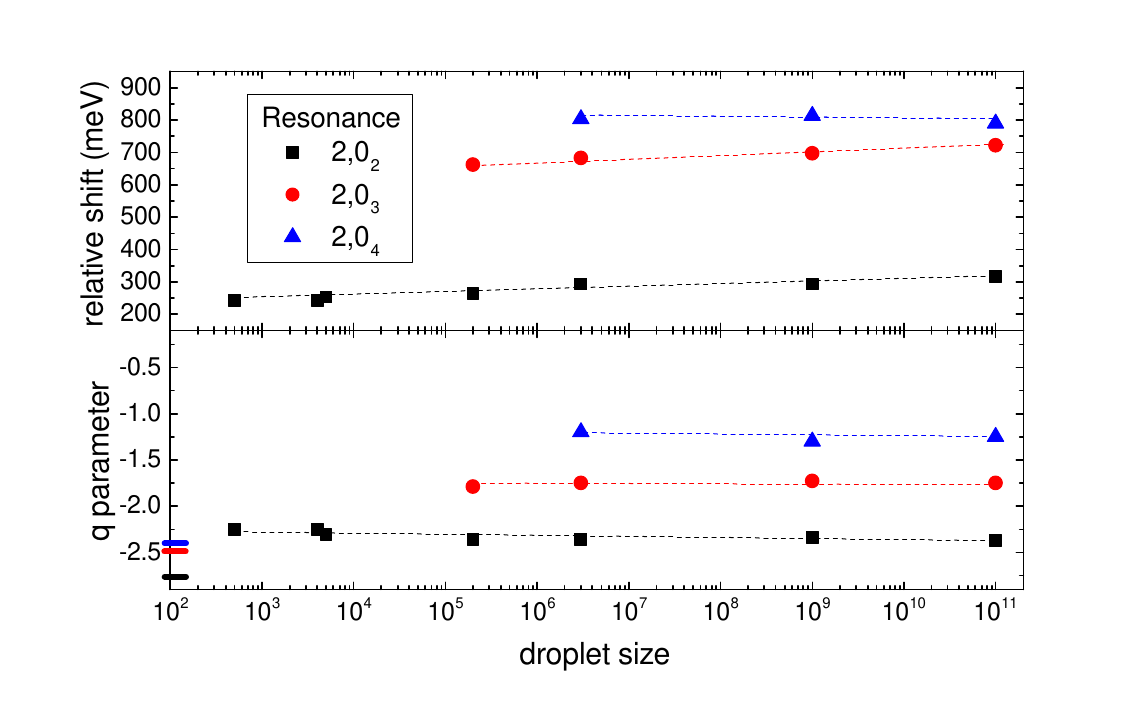}}
\caption{Droplet-size dependence of the shift of the droplet resonance with respect to the atomic resonance (a) and the Fano $q$ parameter (b) for the first three resonances of the 2,0$_n$ series. Dashed lines are linear fits to guide the eyes. The respective atomic q parameters are shown as colored ticks on the y-axis.}
\label{fig4}
\end{center}
\end{figure}

Finally Fig.\,\ref{fig4} shows a) the shifts of the droplet resonances with respect to the atomic resonance, and b), the values of the $q$ parameter for the first three resonances of the 2,0$_{n}$ series as a function of the droplet size. The resonances were fitted using Eq. 1. All three resonances have widths close to that of the droplet 2,0$_{2}$ resonance (400 (50) meV), however their respective $q$ parameters drop to -1.75 for the droplet 2,0$_{3}$ resonance and -1.25 for the droplet 2,0$_{4}$ resonance. The 2,0$_{2}$ resonance shows a slight droplet-size dependence of the relative shift, whereby smaller droplets are closer to the atomic resonance. The higher resonances show no such dependence within the limited statistics of the weak signals. The $q$ parameter appears to be constant over the entire temperature range for all resonances suggesting no change in the relative scattering amplitudes (e.g. excitation, ionization, and autoionization) with droplet size. 

In conclusion, we have observed autoionizing doubly excited resonances in the van der Waals bonded He nanodroplets along with corresponding Fano profiles when scanning over the resonance energies. Relative to the atomic lines, the droplet resonances are significantly broadened and blue shifted similar to features seen in singly excited droplets~\cite{Joppien1993} due to the localized doubly excited He atoms being perturbed by the neighboring droplet environment. Furthermore, a clear size dependence is observed in the overall electron intensity such that the strongest Fano profiles occur for droplets consisting of 10$^{6}$ - 10$^{11}$ atoms; however, the Fano parameters show little to no droplet-size dependence leading one to conclude that the ionization/excitation mechanisms do not depend on droplet size.

Overall, the observation of localized doubly excited states in weakly bound systems opens a variety of new directions for research. For instance, with mass-selected oligomer beam production~\cite{Schoellkopf1994}, one could resolve the molecular states theoretically shown in Fig.\,\ref{fig3} along with possible blueshifts and line broadening for more molecular systems (e.g. dimers and trimers). Additionally, the presence of neighboring atoms allows for competing decay mechanisms such as ICD where instead of energy exchange between excited electrons, the energy of one or both electrons is transferred to a neighboring atom resulting in ionization. The influence of additional decay mechanisms could help to identify the extent of localization of the doubly excited state.

\end{document}